\documentclass[11pt]{article}
\usepackage{amssymb}
\usepackage{amsfonts}
\usepackage{graphicx}
\usepackage{psfrag}
\usepackage{epsfig}
\usepackage{bbm}

\newcommand{\bcen}{\begin{center}}
\newcommand{\ecen}{\end{center}}
\newcommand{\brig}{\begin{flushright}}
\newcommand{\erig}{\end{flushright}}

\def\db#1{\bar D_{#1}}
\def\zb#1{\bar Z_{#1}}
\def\d#1{D_{#1}}
\def\tld#1{\tilde {#1}}
\def\slh#1{\rlap / {#1}}

\newcommand{\txred}{}
\newcommand{\txblu}{}

\newcommand{\txpur}{}

\newcommand{\barray}{\begin{array}}
\newcommand{\eeqarray}{\end{eqnarray*}}
\newcommand{\beqarray}{\begin{eqnarray*}}
\newcommand{\earray}{\end{array}}
\newcommand{\bite}{\begin{itemize}}
\newcommand{\eite}{\end{itemize}}
\newcommand{\bmath}{\begin{displaymath}}
\newcommand{\emath}{\end{displaymath}}
\newcommand{\beq}{\begin{equation}}
\newcommand{\eeq}{\end{equation}}
\newcommand{\bea}{\begin{eqnarray}}
\newcommand{\eea}{\end{eqnarray}}
\newcommand{\bdm}{\begin{displaymath}}
\newcommand{\edm}{\end{displaymath}}
\newcommand{\bqa}{\begin{eqnarray}}
\newcommand{\eqa}{\end{eqnarray}}
\newcommand{\nl}{\nonumber \\}
\newcommand{\bd}{\begin{displaymath}}
\newcommand{\ed}{\end{displaymath}}

\begin{document}

\pagestyle{empty}

\vspace*{13mm}
 
\begin{center}
{\LARGE{\bf A new method for the numerical evaluation of one-loop amplitudes
}}
\vspace*{4mm}
\end{center}

\smallskip
\begin{center}
{\Large{Giovanni Ossola}}  \\
\vspace{8mm}
{\sl Institute of Nuclear Physics, NCSR ``Demokritos'', 
15310 Athens, Greece.
}\\
\vspace{1mm}
{\tt ossola@inp.demokritos.gr}
\vspace*{24mm}

\abstract{
We recently presented a new method for the evaluation of one-loop amplitude of arbitrary scattering processes, in which the 
reduction to scalar integrals is performed at the integrand level. In this talk, we review the main features of the method and briefly summarize the results of the first calculations performed using it.
}
\vspace*{28mm}
\end{center}

\bcen
\emph{To appear in the proceedings of the \\
19th Conference on High Energy Physics (IFAE 2007)\\
11-13 Apr 2004, Napoli, Italy}

\end{center}
\newpage

\section{Introduction and General Motivations}

The experimental programs of LHC require high precision predictions 
for multi-particle processes.
At the tree level, the introduction of efficient recursive 
algorithms \cite{recursive}
improved the theoretical description of such processes.
However, the current need for precision goes beyond tree order. 
The search and the interpretation of new physics requires a precise 
understanding of the Standard Model. We need accurate predictions and 
reliable error estimates. For future experiments, starting with LHC, 
all analyses will require at least next-to-leading order calculations (NLO).

In the last few years, several groups have been working on the 
problem of constructing efficient and automatized methods for the 
computation of one-loop corrections for multi-particle processes.

The standard reference for any one-loop calculation is the fundamental work
of 't~Hooft and Veltman\cite{thv}, and Passarino and Veltman\cite{pv}, 
which already contains many of the ingredients needed to accomplish 
such calculations.
However, after almost three decades, only few one-loop 
calculations involving more 
than five particles have been completed \cite{6part}.

The difficulties arising in this kind of calculations are well known: on the
one 
hand the presence of a very large number of Feynman diagrams, on the other 
the appearance of numerical instabilities (i.e Gram determinants), that 
should be cured or avoided.

Many different interesting techniques have been proposed for NLO calculations:
numerical methods \cite{numerical}, in which tensor integrals are directly
computed numerically;
semi-numerical methods \cite{semi}, in which a reduction to a basis of
known integrals is performed, dealing carefully with 
spurious singularities; 
analytic approaches \cite{ana}, that make use of unitarity cuts to build NLO amplitudes by gluing on-shell tree amplitudes. Some of these techniques require additional rational terms to be computed separately \cite{rats}. Recent complete reviews of existing methods can be found, for example, in Refs.~\cite{review1,review2}.

The main purpose of this talk is to illustrate the new method of reduction 
at the integrand level (OPP reduction) that we developed during the last 
year \cite{opp1, opp2}. The method benefits from previous work of 
Pittau and del Aguila \cite{roots}. In Section~2, we introduce the main features of the method. In Section~3, we deal with the computation of rational terms. Finally, in Section~4, we summarize the applications of the method that have already been implemented.

\section[OPP Reduction]{OPP Reduction}  


Any $m$-point one-loop amplitude can be written, before integration, as
\bqa \label{integrand}
A(\bar q)= \frac{N(q)}{\db{0}\db{1}\cdots \db{m-1}}\, 
\eqa
with $\db{i} = ({\bar q} + p_i)^2-m_i^2$.
The bar denotes objects living in $n=~4+\epsilon$  dimensions and a tilde objects of dimension $\epsilon$. Physical external momenta $p_i$ 
are 4-dimensional objects, while the 
integration momentum $q$ is in general n-dimensional. 
Following this notation, we have
$\bar q^2= q^2+ \tld{q}^2$ and $\db{i} = \d{i} + \tld{q}^2$.

Assuming for the moment that the numerator $N(q)$ is fully 4-dimensional, 
we can rewrite it at the integrand level in terms of $\d{i}$ as
\bqa \label{eq1}
N(q) &=&
\sum_{i_0 < i_1 < i_2 < i_3}^{m-1}
\left[
          \txblu{d( i_0 i_1 i_2 i_3 )} +
     \txred{\tld{d}(q;i_0 i_1 i_2 i_3)}
\right]
\prod_{i \ne i_0, i_1, i_2, i_3}^{m-1} \d{i} \nl
     &+&
\sum_{i_0 < i_1 < i_2 }^{m-1}
\left[
          \txblu{c( i_0 i_1 i_2)} +
     \txred{\tld{c}(q;i_0 i_1 i_2)}
\right]
\prod_{i \ne i_0, i_1, i_2}^{m-1} \d{i} \nl
     &+&
\sum_{i_0 < i_1 }^{m-1}
\left[
          \txblu{b(i_0 i_1)} +
     \txred{\tld{b}(q;i_0 i_1)}
\right]
\prod_{i \ne i_0, i_1}^{m-1} \d{i} \nl
     &+&
\sum_{i_0}^{m-1}
\left[
          \txblu{a(i_0)} +
     \txred{\tld{a}(q;i_0)}
\right]
\prod_{i \ne i_0}^{m-1} \d{i} 
\eqa 

The quantities $d( i_0 i_1 i_2 i_3 )$ are the coefficients of 4-point 
scalar functions with 
denominators labeled by $i_0$, $i_1$, $i_2$, and $i_3$.
In the same way, we call  $c( i_0 i_1 i_2)$, $b( i_0 i_1)$, $a( i_0)$
the coefficients of all possible 3-point, 2-point and 1-point scalar functions, respectively.

The quantities $\tld{d}$, $\tld{c}$, $\tld{b}$, $\tld{a}$ are what we define ``spurious'' terms, i.e. terms that are present in the decomposition at the integrand level, but they will vanish upon integration. These terms still depend on the integration momentum $q$.

In order to perform the reduction, we should first find a general
explicit expression for the spurious term.
Following Ref.~\cite{roots}, we rewrite all $q$'s in $N(q)$ by means of
 a basis of massless vectors $\ell_i^\mu$
\beq
q^{\mu} = -p_0^{\mu} + \sum_{i=1}^{4} G_i\, \ell_i^\mu
\quad \quad  [{\ell_i}^2=0]
\eeq
where
\bd
k_1   = \ell_1 + \alpha_1
\ell_2\,,~~~k_2   = \ell_2 + \alpha_2 \ell_1 \,,\,\,
 k_i = p_i-p_0
\ed
and
\bd
\txpur{\ell_3}^\mu = <\ell_1| \gamma^\mu | \ell_2]\,,~
\txpur{\ell_4}^\mu = <\ell_2| \gamma^\mu | \ell_1]\,\, .
\ed
The coefficients $G_i$ either reconstruct denominators $\d{i}$
or vanish upon integration.
In the first case, they give rise to $d$,  $c$,  
$b$,  $a$ coefficients, while in the second one
they form the spurious $\tld{d}$,  $\tld{c}$,
$\tld{b}$,  $\tld{a}$ coefficients. 

As an example, we explicitly report the expressions of the $\tld{d}$(q)
term:
\bd
\txred{\tld{d}(q)} = \txred{\tld{d}}\, 
Tr[(\slh{q}+\slh{p_0}) \slh{\ell_1} \slh{\ell_2} \slh{k_3}
\gamma_5] \,.
\ed
No other $q$-dependent structure can appear at the level of 4-point functions.
For the 3-point functions, there are six possible terms that can contribute 
to $\tld{c}$(q):
\bd
\txred{\tld{c}(q)}= \sum_{j=1}^{j_{max}} \left\{
 \txred{\tld{c}_{1j}}[(q+p_0)\cdot \ell_3]^j
+\txred{\tld{c}_{2j}}[(q+p_0)\cdot \ell_4]^j
        \right\} \,\, .
\ed
In the renormalizable gauge, $j_{max}= 3$.
The 2- and 1-point spurious functions $\tld{b}$(q) and $\tld{a}$(q) 
contain eight and four terms, respectively \cite{opp1}.

After fixing the form of all the spurious terms,
our calculation is reduced to the algebraic problem of
extracting all the coefficients. This is achieved simply
by evaluating N(q) at different values of the integration momentum $q$.
As a further simplification, there is a very good set of such points: we can 
use values of $q$ for which a subset of denominators $\d{i}$ vanish. 
Operating in this manner, the system becomes ``triangular''; 
we can solve first for 4-point functions, then 3-point functions and so on.

To conclude this section, let us summarize the recipe for the calculation. 
We first calculate all the coefficients in Eq.~(\ref{eq1}), by evaluating the
numerator of the integrand $N(q)$ for a set of values of the 
integration momentum $q$.
Note that we do not need to repeat this for all Feynman diagrams:
we can group them and expand for (sub)amplitudes directly. We just 
need to specify external momenta, polarization vectors and masses and
proceed with the reduction.
Concerning  $N(q)$ we can choose how to proceed according 
to the specific calculation: as an interesting development,
$N(q)$ could be determined numerically via recursion relations.
Once all the coefficients have been determined, we can multiply
them by the corresponding scalar integrals.
For massive integrals we use FF by G.~J.~van Oldenborgh \cite{ff},
while in the massless case the integrals are evaluated using the code
OneLOop by A.~van~Hameren \cite{avh}.

\section{Computation of rational terms}

As mentioned in the previous section, the reduction has been performed 
assuming a purely four dimensional numerator (this singles out 
the so called cut-constructable part of the amplitude).
In this section we describe our method to calculate the rational 
parts of the amplitude.
In our approach, these contributions originate from the fact that up to
now we expressed the numerator in terms of  $\d{i}$, while the functions 
appearing in the denominator are n-dimensional  $\db{i}$.
Let us go back to the integrand  $A(\bar q)$ of Eq.~(\ref{integrand})
and insert the expression for $N(q)$ of Eq.~(\ref{eq1}), that we 
obtained after determining
all the coefficients for both for regular and spurious terms.
Now, rewriting $\d{i}$ by means of
\bd
\frac{\d{i}}{\db{i}} = \txpur{\zb{i}}\,,~~~~{\rm with}~~~
\txpur{\zb{i}\equiv \left(1- \frac{\tld{q}^2}{\db{i}} \right)}
\ed
 we obtain
\bqa
A(\bar q) &=&
\sum_{i_0 < i_1 < i_2 < i_3}^{m-1}
\frac{
          d( i_0 i_1 i_2 i_3 ) +
     \tld{d}(q;i_0 i_1 i_2 i_3)
}{\db{i_0} \db{i_1} \db{i_2} \db{i_3}}
\txpur{\prod_{i \ne i_0, i_1, i_2, i_3}^{m-1} \zb{i}} \nl
     &+&
\sum_{i_0 < i_1 < i_2 }^{m-1}
\frac{
          c( i_0 i_1 i_2) +
     \tld{c}(q;i_0 i_1 i_2)
}{\db{i_0} \db{i_1} \db{i_2}}
\txpur{\prod_{i \ne i_0, i_1, i_2}^{m-1} \zb{i}} \nl
     &+&
\sum_{i_0 < i_1 }^{m-1}
\frac{
          b(i_0 i_1) +
     \tld{b}(q;i_0 i_1)
}{\db{i_0} \db{i_1}}
\txpur{\prod_{i \ne i_0, i_1}^{m-1} \zb{i} }\nl
     &+&
\sum_{i_0}^{m-1}
\frac{
          a(i_0) +
     \tld{a}(q;i_0)
}{\db{i_0}}\txpur{ \prod_{i \ne i_0}^{m-1} \zb{i}} \,\, .
\eqa  
The rational part is produced, after integrating over $d^n q$, by the
$\tld{q}^2$ dependence in $\zb{i}$. 
The expressions for all relevant integrals are reported 
in the Appendix of Ref.~\cite{opp2}. 

In addition to the one just described, 
there might be other 
sources of rational terms coming from objects of dimension $\epsilon$ in
the numerator N(q), according to the specific calculation.
These, for example, could originate
from the contraction of Dirac matrices or from 
powers of $\bar q^2$ \cite{pittau2}.


\section{Numerical Tests}


As an example of applications of the method, we calculated 4-photon and 
6-photon amplitudes, via fermionic loop of mass $m_f$ \cite{opp2}.
Those processes do not have an immediate physical interest,  
however they provide a playground
for testing and comparing different methods of calculation.

In order to perform the calculation, for each phase-space point we should 
provide as input parameters: the external momenta $p_i$ (for this particular example they are massless, i.e. $p_i^2=0$); the masses of propagators in the loop (fermion masses $m_f$ in this example); the polarization vectors for the photons: here we use various helicity configurations.

For the 4-photon amplitudes, we performed a comparison with the 
analytic result presented by Gounaris et al. \cite{gou}.
We compared separately the rational part and the cut-constructable part, both
in the massive and massless cases, finding perfect agreement.
We repeated the calculation for the three helicity configurations $F^f_{++++}$,
$F^f_{+++-}$ and $F^f_{++--}$ presented in the Appendix of Ref.~\cite{gou}.

Coming to the 6-photons, there are few results available in the literature for the massless case. 
Some time ago, Mahlon presented an exact analytic result for the helicity 
configuration $[++----]$ \cite{mahlon}. More recently, Nagy and Soper, with a fully numerical approach, obtained the results
for the configurations $[++----]$ and $[++--+-]$ \cite{nagy}. 
The same results were also recently presented by Binoth et al. \cite{binoth}, that also provide analytic expressions.
Our results are in full agreement with all previous calculations.
We also checked that the cut-constructable part for configurations $[+-----]$ and $[------]$ and the rational terms for all helicity configurations are identically zero.

Finally we calculated the 6-photon amplitudes with massive internal fermions.
Also in this case there is no contribution from rational parts. The results have been presented 
in Ref.~\cite{opp2}.

\section{Summary and Conclusions}

The discovery potential of LHC requires NLO calculations. 
At present, there is a variety of interesting options available to perform
 one-loop multi-leg calculations, however not a universal method.

We recently proposed a method for the numerical evaluation of one-loop amplitudes in which the reduction is performed at the integrand level.
The method is rather ``young'' and there is still room 
for  improvements and optimizations, that are currently in progress. 
The first results are encouraging. 
We plan to present more results soon, for processes of interest 
at future experiments.
  

\bigskip
\bigskip

\noindent {\bf Acknowledgments} \\ 
Work done in collaboration with Costas Papadopoulos and Roberto Pittau. Many thanks to Andre van Hameren for careful numerical comparisons, and to Pierpaolo Mastrolia and Zoltan Nagy for fruitful communications. This research was supported by the ToK Program ``ALGOTOOLS'' (MTKD-CT-2004-014319).

\end{document}